\documentclass{article}

\newcommand{\beq}[1]{\begin{equation}\label{#1}}
\newcommand{\eeq}{\end{equation}}
\newcommand{\bear}[1]{\begin{eqnarray}\label{#1}}
\newcommand{\ear}{\end{eqnarray}}
\newcommand{\nn}{\nonumber}
\newcommand{\rf}[1]{(\ref{#1})}

\catcode`\@=11 \@addtoreset{equation}{section}\catcode`\@=12

\def\barr{\left(\begin{array}}
\def\earr{\end{array}\right)}
\def\beq#1{\begin{equation}\label{#1}}
\def\eeq{\end{equation}}
\def\ber#1{\begin{eqnarray}\label{#1} }
\def\eer{\end{eqnarray}}

\newcommand{\N}{ {\bf N} }
\newcommand{\R}{ {\bf R} }

\newcommand{\sign}{ \mbox{\rm sign} }
\newcommand{\e}{ \mbox{\rm e} }
\newcommand{\eps}{ \varepsilon }

\newcommand{\p}{\partial}

\newcommand{\tri}{\Delta}

\newcommand{\im}{{\rm i}}
\newcommand{\fnm}{\footnotemark}
\newcommand{\fnt}{\footnotetext}

\begin{document}

 \begin{center}
 \large \bf
 On exact solutions in multidimensional gravity
   with  antisymmetric forms

 \end{center}

 \vspace{0.3truecm}

 \begin{center}

 \normalsize\bf

 V. D. Ivashchuk\fnm[1]\fnt[1]{e-mail: ivas@rgs.phys.msu.su}

 \it
  Center for Gravitation and Fundamental Metrology,
  VNIIMS, 3-1 M. Ulyanovoy Str., Moscow, 119313, Russia and

 \it Institute of Gravitation and Cosmology,
  Peoples' Friendship University of Russia,
  6 Miklukho-Maklaya St., Moscow 117198, Russia

 \end{center}

 \begin{abstract}

 This short review deals with a multidimensional gravitational model
 containing  dilatonic scalar fields
 and antisymmetric forms. The manifold
 is chosen in the product form. The sigma-model approach and
 exact solutions are reviewed.

 \end{abstract}


  \section{Introduction}

 In these lectures we consider  several classes of the
 exact solutions for the multidimensional gravitational
 model governed by the  action
 \bear{2.1}
  S_{act} =&& \frac{1}{2\kappa^{2}}
  \int_{M} d^{D}z \sqrt{|g|} \{ {R}[g] - 2 \Lambda - h_{\alpha\beta}
  g^{MN} \partial_{M} \varphi^\alpha \partial_{N} \varphi^\beta
  \\ \nn
  && - \sum_{a \in \Delta}
  \frac{\theta_a}{n_a!} \exp[ 2 \lambda_{a} (\varphi) ] (F^a)^2_g \}
   + S_{GH},
 \ear
where $g = g_{MN} dz^{M} \otimes dz^{N}$ is the metric
on the manifold $M$, ${\dim M} = D$, $\varphi=(\varphi^\alpha)\in \R^l$
is a vector from dilatonic scalar fields,
 $(h_{\alpha\beta})$ is a non-degenerate symmetric
 $l\times l$ matrix ($l\in \N$),
 $\theta_a  \neq 0$,
 $$F^a =  dA^a
  =\frac{1}{n_a!} F^a_{M_1 \ldots M_{n_a}}
  dz^{M_1} \wedge \ldots \wedge dz^{M_{n_a}}
 $$
is a $n_a$-form ($n_a \geq 2$) on a $D$-dimensional manifold $M$,
 $\Lambda$ is a cosmological constant
and $\lambda_{a}$ is a $1$-form on $\R^l$ :
 $\lambda_{a} (\varphi) =\lambda_{a \alpha}\varphi^\alpha$,
 $a \in \Delta$, $\alpha=1,\ldots,l$.
In (\ref{2.1})
we denote $|g| = |\det (g_{MN})|$,
  $(F^a)^2_g =
   F^a_{M_1 \ldots M_{n_a}} F^a_{N_1 \ldots N_{n_a}}
   g^{M_1 N_1} \ldots g^{M_{n_a} N_{n_a}},$
 $a \in \Delta$, where $\Delta$ is some finite set, and $S_{\rm GH}$ is the
standard (Gibbons-Hawking) boundary term. In the models
with one time all $\theta_a =  1$  when the signature of the metric
is $(-1,+1, \ldots, +1)$.

For certain field  contents with
distinguished values of total dimension $D$, ranks $n_a$,
dilatonic couplings $\lambda_{a}$  and $\Lambda = 0$
such Lagrangians appear as ``truncated'' bosonic sectors
(i.e. without Chern-Simons terms) of certain
supergravitational theories or low-energy limit of superstring models
 (see \cite{St,IMtop} and references therein).

 \section{The model}

\subsection{Ansatz for composite  p-branes }

Let us consider the manifold
 \beq{2.10}
  M = M_{0}  \times M_{1} \times \ldots \times M_{n},
 \eeq
with the metric
 \beq{2.11}
  g= e^{2{\gamma}(x)} \hat{g}^0  +
  \sum_{i=1}^{n} e^{2\phi^i(x)} \hat{g}^i ,
 \eeq
where $g^0  = g^0 _{\mu \nu}(x) dx^{\mu} \otimes dx^{\nu}$
is an arbitrary metric with any signature on the manifold $M_{0}$
and $g^i  = g^{i}_{m_{i} n_{i}}(y_i) dy_i^{m_{i}} \otimes dy_i^{n_{i}}$
is a metric on $M_{i}$  satisfying the equation
 \beq{2.13}
  R_{m_{i}n_{i}}[g^i ] = \xi_{i} g^i_{m_{i}n_{i}},
 \eeq
 $m_{i},n_{i}=1, \ldots, d_{i}$; $\xi_{i}= {\rm const}$,
 $i=1,\ldots,n$. Here $\hat{g}^{i} = p_{i}^{*} g^{i}$ is the
pullback of the metric $g^{i}$  to the manifold  $M$ by the
canonical projection: $p_{i} : M \rightarrow  M_{i}$,
 $i = 0,\ldots, n$. Thus, $(M_i, g^i )$  are Einstein spaces,
 $i = 1,\ldots, n$.
The functions $\gamma, \phi^{i} : M_0 \rightarrow \R $ are smooth.
We denote $d_{\nu} = {\rm dim} M_{\nu}$; $\nu = 0, \ldots, n$;
 $D = \sum_{\nu = 0}^{n} d_{\nu}$.
We put any manifold $M_{\nu}$, $\nu = 0,\ldots, n$,
to be oriented and connected.
Then the volume $d_i$-form
 \beq{2.14}
  \tau_i  \equiv \sqrt{|g^i(y_i)|}
   \ dy_i^{1} \wedge \ldots \wedge dy_i^{d_i},
 \eeq
and signature parameter
 \beq{2.15}
  \varepsilon(i)  \equiv {\rm sign}( \det (g^i_{m_i n_i})) = \pm 1
 \eeq
are correctly defined for all $i=1,\ldots,n$.

Let $\Omega = \Omega(n)$  be a set of all non-empty
subsets of $\{ 1, \ldots,n \}$.
The number of elements in $\Omega$ is $|\Omega| = 2^n - 1$.
For any $I = \{ i_1, \ldots, i_k \} \in \Omega$, $i_1 < \ldots < i_k$,
we denote
 \bear{2.16}
  \tau(I) \equiv \hat{\tau}_{i_1}  \wedge \ldots \wedge \hat{\tau}_{i_k},
  \\
  \label{2.17}
  \eps(I) \equiv \eps(i_1) \ldots \eps(i_k),  \\
  \label{2.19}
   d(I) \equiv  \sum_{i \in I} d_i.
 \ear
Here $\hat{\tau}_{i} = p_{i}^{*} \hat{\tau}_{i}$ is the
pullback of the form $\tau_i$  to the manifold  $M$ by the
canonical projection: $p_{i} : M \rightarrow  M_{i}$,
 $i = 1,\ldots, n$. We also put $\tau(\emptyset)= \eps(\emptyset)=
 1$ and $d(\emptyset)=0$.

For fields of forms we consider the following composite electromagnetic
ansatz
 \ber{2.1.1}
  F^a=\sum_{I\in\Omega_{a,e}}{\cal F}^{(a,e,I)}+
  \sum_{J\in\Omega_{a,m}}{\cal F}^{(a,m,J)}
 \eer
where
 \bear{2.1.2}
  {\cal F}^{(a,e,I)}=d\Phi^{(a,e,I)}\wedge\tau(I), \\
  \label{2.1.3}
  {\cal F}^{(a,m,J)}= e^{-2\lambda_a(\varphi)}*(d\Phi^{(a,m,J)}
  \wedge\tau(J))
 \ear
 are elementary forms of electric and magnetic types respectively,
 $a\in\tri$, $I\in\Omega_{a,e}$, $J\in\Omega_{a,m}$ and
 $\Omega_{a,v} \subset \Omega$, $v = e,m$. In (\ref{2.1.3})
 $*=*[g]$ is the Hodge operator on $(M,g)$.
For scalar functions we put
 \ber{2.1.5}
   \varphi^\alpha=\varphi^\alpha(x), \quad
   \Phi^s=\Phi^s(x),
 \eer
 $s\in S$. Thus $\varphi^{\alpha}$ and $\Phi^s$ are functions on $M_0$.

Here and below
 \ber{2.1.6}
  S=S_e \sqcup S_m, \quad
  S_v=\sqcup_{a\in\tri}\{a\}\times\{v\}\times\Omega_{a,v},
 \eer
 $v=e,m$. Here and in what follows $\sqcup$ means the union
of non-intersecting sets.
The set $S$ consists of elements $s=(a_s,v_s,I_s)$,
where $a_s \in \tri$ is colour index, $v_s = e, m$ is electro-magnetic
index and set $I_s \in \Omega_{a_s,v_s}$ describes the location
of brane.

Due to (\ref{2.1.2}) and (\ref{2.1.3})
 \ber{2.1.7}
  d(I)=n_a-1, \quad d(J)=D-n_a-1,
 \eer
for $I\in\Omega_{a,e}$ and $J\in\Omega_{a,m}$ (i.e. in electric
and magnetic case, respectively). The sum of worldvolume dimensions
for electric and magnetic branes corresponding to the same form is equal
to $D-2$, it does not depend upon the rank of the form.

 \subsection{The sigma model}

Let $d_0 \neq 2$ and
 \ber{2.2.1}
   \gamma=\gamma_0(\phi) \equiv
    \frac1{2-d_0}\sum_{j=1}^nd_j\phi^j,
 \eer
i.e. the generalized harmonic gauge (frame) is used.
As we shall see below the equations of motions have a
rather simple form in this gauge.

 \subsubsection{Restrictions on $p$-brane configurations.}

Here we present two restrictions on the sets of $p$-branes
that guarantee  the block-diagonal form of the  energy-momentum tensor
and the existence of the sigma-model representation (without additional
constraints) \cite{IMC} (see also \cite{AR}).

 {\bf Restriction 1} reads
  \beq{2.2.2a}
  {\bf (R1)} \quad d(I \cap J) \leq d(I)  - 2,
  \eeq
for any $I,J \in\Omega_{a,v}$, $a\in\tri$, $v= e,m$ (here $d(I) = d(J)$).

 {\bf Restriction 2} has the following form
 \beq{2.2.3a}
  {\bf (R2)} \quad d(I \cap J) \neq 0 \ for \ d_0 = 1, \qquad
   d(I \cap J) \neq 1  \quad for \ d_0 = 3
  \eeq
(see (\ref{2.1.7})).

 \subsubsection{Sigma-model action for harmonic gauge}

It was proved in \cite{IMC} that equations of motion for the model
 (\ref{2.1}) and the Bianchi identities:
   $ d{\cal F}^s=0$,
   $s \in S_m$, for fields from (\ref{2.11}),
  (\ref{2.1.1})--(\ref{2.1.5}), when Restrictions 1 and 2
are  imposed, are equivalent to equations of motion for the $\sigma$-model
governed by the action
 \bear{2.2.7}
  S_{\sigma 0} = && \frac{1}{2 \kappa_0^2}
  \int d^{d_0}x\sqrt{|g^0|}\biggl\{R[g^0]-\hat G_{AB}
  g^{0\mu\nu}\p_\mu\sigma^A\p_\nu\sigma^B  \\ \nn
  &&-\sum_{s\in S}\eps_s \exp{(-2U_A^s\sigma^A)}
  g^{0\mu\nu} \p_\mu\Phi^s\p_\nu\Phi^s - 2V \biggr\},
 \ear
where $(\sigma^A)=(\phi^i,\varphi^\alpha)$, $k_0 \neq 0$, the index set
 $S$ is defined in (\ref{2.1.6}),
 \beq{2.2.8}
  V = {V}(\phi)
  = \Lambda e^{2 {\gamma_0}(\phi)}
  -\frac{1}{2}   \sum_{i =1}^{n} \xi_i d_i e^{-2 \phi^i
  + 2 {\gamma_0}(\phi)}
 \eeq
is the potential,
 \ber{2.2.9}
  (\hat G_{AB})=\barr{cc}
  G_{ij}& 0\\
  0& h_{\alpha\beta}
 \earr,
 \eer
is the target space metric
with
  \ber{2.2.10}
   G_{ij}= d_i \delta_{ij}+\frac{d_i d_j}{d_0-2},
  \eer
and co-vectors
 \ber{2.2.11}
  U_A^s =
  U_A^s \sigma^A = \sum_{i \in I_s} d_i \phi^i -
  \chi_s \lambda_{a_s}(\varphi),
  \quad
  (U_A^s) =  (d_i \delta_{iI_s}, -\chi_s \lambda_{a_s \alpha}),
 \eer
 $s=(a_s,v_s,I_s)$.
Here $\chi_e=+1$ and $\chi_m=-1$;
 \ber{2.2.12}
  \delta_{iI}=\sum_{j\in I}\delta_{ij}
 \eer
is an indicator of $i$ belonging
to $I$: $\delta_{iI}=1$ for $i\in I$ and $\delta_{iI}=0$ otherwise; and
 \bear{2.2.13}
   \eps_s=(-\eps[g])^{(1-\chi_s)/2}\eps(I_s) \theta_{a_s},
 \ear
 $s\in S$, $\eps[g]\equiv\sign\det(g_{MN})$. More explicitly
 (\ref{2.2.13}) reads
 \beq{2.2.13a}
   \eps_s=\eps(I_s) \theta_{a_s} \ {\rm for} \ v_s = e; \qquad
   \eps_s = -\eps[g] \eps(I_s) \theta_{a_s}, \ {\rm for} \ v_s = m.
 \eeq

For finite internal space volumes $V_i$ (e.g. compact $M_i$)
and electric $p$-branes
the action (\ref{2.2.7}) coincides with the action (\ref{2.1}) when
 $\kappa^{2} = \kappa^{2}_0 \prod_{i=1}^{n} V_i$.

 {\bf Sigma-model  with constraints.}
In \cite{IMC} a general proposition concerning the sigma-model
representation when the {\bf Restrictions 1} and {\bf 2} are removed
is presented.  In this case the stress-energy tensor
is not identically block-diagonal and  several additional constraints
on the field configurations appear \cite{IMC}.

We note that the symmetries of target space metric were studied in \cite{Iv3}.
Sigma-model approach for models  with non-diagonal metrics
was suggested in  \cite{GR}.

 \section{Solutions governed by harmonic functions}

 \subsection{Solutions with  block-orthogonal $U^s$}

Here we consider a special class of solutions to equations of motion
governed by several harmonic functions when all factor spaces are
Ricci-flat and cosmological constant is zero, i.e.
  $\xi_i = \Lambda = 0$,  ($i = 1,\ldots,n$) \cite{IMBl,GrI}.
In certain situations these solutions describe extremal $p$-brane black holes
charged by fields of forms.

The solutions crucially depend upon  scalar products
of $U^s$-vectors $(U^s,U^{s'})$; $s,s' \in S$, where
 \ber{3.1.1}
  (U,U')=\hat G^{AB} U_A U'_B,
 \eer
for $U = (U_A), U' = (U'_A) \in \R^N$, $N = n + l$ and
 \beq{3.1.2}
  (\hat G^{AB})=\left(\begin{array}{cc}
  G^{ij}&0\\
  0&h^{\alpha\beta}
  \end{array}\right)
 \eeq
is matrix inverse to  the matrix
 (\ref{2.2.9}). Here (as in \cite{IMZ})
 \beq{3.1.3}
    G^{ij}=\frac{\delta^{ij}}{d_i}+\frac1{2-D},
 \eeq
 $i,j=1,\dots,n$.

The scalar products (\ref{3.1.1}) for vectors
 $U^s$  were calculated in \cite{IMC}
 \ber{3.1.4}
  (U^s,U^{s'})=d(I_s\cap I_{s'})+\frac{d(I_s)d(I_{s'})}{2-D}+
  \chi_s\chi_{s'}\lambda_{a_s \alpha} \lambda_{a_{s'}
  \beta} h^{\alpha \beta},
 \eer
where $(h^{\alpha\beta})=(h_{\alpha\beta})^{-1}$; and  $s=(a_s,v_s,I_s)$,
 $s'=(a_{s'},v_{s'},I_{s'})$ belong to $S$. This relation is a
very important one since it encodes  $p$-brane data
(e.g. intersections) in
scalar products of $U$-vectors.

Let
 \beq{3.1.5}
  S=S_1\sqcup\dots\sqcup S_k,
 \eeq
 $S_i\ne\emptyset$, $i=1,\dots,k$, and
 \beq{3.1.6}
  (U^s,U^{s'})=0
 \eeq
for all $s\in S_i$, $s'\in S_j$, $i\ne j$; $i,j=1,\dots,k$. Relation
 (\ref{3.1.5}) means that the set $S$ is a union of $k$ non-intersecting
(non-empty) subsets $S_1,\dots,S_k$.

Here we consider exact solutions in the
model (\ref{2.1}), when vectors $(U^s,s\in S)$ obey the block-orthogonal
decomposition (\ref{3.1.5}), (\ref{3.1.6}) with scalar products defined in
 (\ref{3.1.4}) \cite{IMBl}. These solutions may be obtained from the
following proposition.

 {\bf Proposition  \cite{IMBl}.} {\em Let $(M_0,g^0)$ be Ricci-flat:
  $R_{\mu\nu}[g^0]=0$.
Then the field configuration
 \ber{3.1.7}
  g^0, \qquad \sigma^A=\sum_{s\in S}\eps_sU^{sA}\nu_s^2\ln H_s, \qquad
  \Phi^s=\frac{\nu_s}{H_s},
\eer
 $s\in S$, satisfies to field equations corresponding to action
(\ref{2.2.7}) with
 $V=0$ if (real) numbers $\nu_s$ obey the relations
 \ber{3.1.8}
   \sum_{s'\in S}(U^s,U^{s'})\eps_{s'}\nu_{s'}^2=-1
 \eer
 $s\in S$, functions $H_s >0$ are harmonic, i.e.
 $\tri[g^0]H_s=0$,
 $s\in S$, and $H_s$ are coinciding inside blocks:
 $H_s=H_{s'}$
for $s,s'\in S_i$, $i=1,\dots,k$.}

Using the sigma-model solution from Proposition
and relations for contravariant components \cite{IMC}:
 \ber{2.2.38}
   U^{si}=\delta_{iI_s}-\frac{d(I_s)}{D-2}, \quad
   U^{s\alpha}=-\chi_s\lambda_{a_s}^\alpha,
 \eer
 $s=(a_s,v_s,I_s)$,  we get \cite{IMBl}:
 \bear{3.1.11}
   g= \left(\prod_{s\in S}H_s^{2d(I_s)\eps_s\nu_s^2}\right)^{1/(2-D)}
   \left\{\hat{g}^0+ \sum_{i=1}^n
   \left(\prod_{s\in S}H_s^{2\eps_s\nu_s^2\delta_{iI_s}}\right)
   \hat{g}^i\right\}, \\
   \label{3.1.14}
   \varphi^\alpha=-\sum_{s\in S}\lambda_{a_s}^\alpha\chi_s
   \eps_s\nu_s^2\ln H_s, \\
    \label{3.1.15}
    F^a=\sum_{s\in S}{\cal F}^s\delta_{a_s}^a,
  \ear
where $i=1,\dots,n$, $\alpha=1,\dots,l$, $a \in \Delta$ and
  \ber{3.1.16}
    {\cal F}^s=\nu_sdH_s^{-1}\wedge\tau(I_s)  \mbox{ for } v_s=e, \\
    \label{3.1.17}
    {\cal F}^s=\nu_s(*_0dH_s)\wedge\tau(\bar I_s) \mbox{ for } v_s=m,
  \eer
 $H_s$ are harmonic functions on $(M_0,g^0)$ coinciding inside blocks
(i.e. $H_s=H_{s'}$ for $s,s'\in S_i$, $i=1,\dots,k$)
and relations  (\ref{3.1.8}) on parameters $\nu_s$ are imposed. Here
the matrix $((U^s,U^{s'}))$ and parameters $\eps_s$, $s\in S$,
are defined in (\ref{3.1.4}) and
 (\ref{2.2.13}), respectively;
 $\lambda_a^\alpha=  h^{\alpha\beta}\lambda_{\beta a}$, $*_0=*[g^0]$ is the
Hodge operator  on $(M_0,g^0)$ and $\bar I$  is defined
as follows
 \ber{2.2.5}
  \bar I \equiv  I_0 \setminus I,  \qquad I_0 = \{1,\ldots,n\}.
 \eer
In (\ref{3.1.17}) we redefined the sign of $\nu_{s}$-parameter
compared to (\ref{2.1.3}).

\subsubsection{Solutions with orthogonal $U^s$ }

Let us consider the orthogonal case \cite{IMC}
 \beq{3.1.23}
   (U^s,U^{s'})=0, \qquad  s \neq s',
 \eeq
 $s, s' \in S$. Then relation (\ref{3.1.8}) reads as follows
 \beq{3.1.24}
  (U^s,U^{s}) \eps_{s} \nu_{s}^2=-1,
 \eeq
 $s \in S$. This implies $(U^s,U^{s}) \neq 0$ and
 \beq{3.1.25}
   \eps_{s} (U^s,U^{s}) < 0,
 \eeq
for all $s \in S$.
For $d(I_s) < D - 2$ and
 $\lambda_{a_s \alpha} \lambda_{a_{s} \beta} h^{\alpha \beta} \geq 0$ we get
from (\ref{3.1.4})
  $(U^s,U^{s}) > 0$,
and, hence, $\eps_{s} < 0$,
  $s \in S$. If $\theta_a > 0$ for all $a \in \Delta$,
then
 \beq{3.1.28}
  \eps(I_s) = -1 \ {\rm for} \ v_s = e; \qquad
  \eps(I_s) = \eps[g] \ {\rm for} \ v_s = m.
 \eeq
For pseudo-Euclidean metric $g$ all $\eps(I_s) = -1$ and, hence,
all $p$-branes should contain time manifold. For the  metric $g$ with
the Euclidean signature only magnetic $p$-branes can exist in this case.

From  scalar products (\ref{3.1.4}) and the
orthogonality condition (\ref{3.1.23}) we get the "orthogonal"
intersection rules \cite{AR,IMC}
 \ber{3.1.30}
  d(I_s\cap I_{s'}) =
  \frac{d(I_s)d(I_{s'})}{D - 2} - \chi_s\chi_{s'}\lambda_{a_s \alpha}
  \lambda_{a_{s'} \beta} h^{\alpha \beta} \equiv
  \Delta(s,s'),
 \eer
for $s=(a_s,v_s,I_s) \neq s'=(a_{s'},v_{s'},I_{s'})$.
(For pure electric case see also \cite{IM11,IM12}.)

Certain  supersymmetric solutions in $D = 11$ supergravity
defined on product of  Ricci-flat spaces were  considered in \cite{Iv2}.

\subsection{General Toda-type solutions with one harmonic function}

It is well known
that geodesics of the target space equipped with some
harmonic function on a three-dimensional space generate a solution
to the $\sigma$-model equations \cite{NK,KSMH}.
Here we apply this null-geodesic method  to our sigma-model and
obtain a  new class of solutions in
multidimensional gravity with  $p$-branes
governed by one harmonic function $H$.

\subsubsection{Toda-like Lagrangian}

Action (\ref{2.2.7}) may be also written in the form
  \beq{3.2.1n}
    S_{\sigma 0} =  \frac{1}{2 \kappa_0^2} \int d^{d_0}x\sqrt{|g^0|}
     \{ R[g^0]-  {\cal G}_{\hat A\hat B}(X)
    g^{0 \mu \nu} \p_\mu X^{\hat A} \p_\nu X^{\hat B}   - 2V \}
  \eeq
where $X = (X^{\hat A})=(\phi^i,\varphi^\alpha,\Phi^s)\in{\bf
       R}^{N}$, and minisupermetric
   ${\cal G}=
   {\cal G}_{\hat A\hat B}(X)dX^{\hat A}\otimes dX^{\hat B}$
on minisuperspace
   ${\cal M}={\bf R}^{N}$,  $N = n+l+|S|$
  ($|S|$ is the number of elements in $S$) is defined by the relation
 \beq{3.2.3n}
   ({\cal G}_{\hat A\hat B}(X))=\left(\begin{array}{ccc}
   G_{ij}&0&0\\[5pt]
          0&h_{\alpha\beta}&0\\[5pt]
        0&0&\eps_s \exp(-2U^s(\sigma))\delta_{ss'}
 \end{array}\right).
 \eeq

Here we consider exact solutions to field equations corresponding
to the action  (\ref{3.2.1n})
 \bear{3.2.4}
  R_{\mu\nu}[g^0]=
  {\cal G}_{\hat A \hat B}(X) \p_{\mu} X^{\hat A} \p_\nu  X^{\hat B}
  + \frac{2V}{d_0-2}g_{\mu\nu}^0,    \\
  \label{3.2.5}
  \frac{1}{\sqrt{|g^0|}} \p_\mu [\sqrt{|g^0|}
  {\cal G}_{\hat C \hat B}(X)g^{0\mu\nu }  \p_\nu  X^{\hat B}]
  - \frac{1}{2} {\cal G}_{\hat A \hat B, \hat C}(X)
  g^{0,\mu \nu} \p_{\mu} X^{\hat A} \p_\nu  X^{\hat B} = V_{,\hat C},
 \ear
 $s\in S$. Here  $V_{,\hat C} = \p V / \p  X^{\hat C}$.

We put
 \ber{3.2.6}
   X^{\hat A}(x) =  F^{\hat A}(H(x)),
 \eer
where $F: (u_{-}, u_{+}) \rightarrow \R^{N}$  is a smooth function,
 $H: M_0 \rightarrow \R $ is a harmonic function on $M_0$ (i.e.
 $\tri[g^0]H=0$), satisfying  $u_{-} < H(x)  < u_{+}$ for all $x \in M_0$.
Let all factor spaces are  Ricci-flat
and cosmological constant is zero, i.e. relation
 $\xi_i = \Lambda = 0$
is satisfied. In this case the potential is zero : $V = 0$.
It may be verified that the field equations (\ref{3.2.4}) and (\ref{3.2.5})
are satisfied identically  if $F = F(u)$ obey the
Lagrange equations  for the Lagrangian
 \beq{3.2.12}
  L =  \frac{1}{2} {\cal G}_{\hat A\hat B}(F) \dot F^{\hat A}
  \dot F^{\hat B}
 \eeq
with the zero-energy constraint
 \beq{3.2.13}
  E =  \frac{1}{2} {\cal G}_{\hat A\hat B}(F) \dot F^{\hat A}
                                              \dot F^{\hat B} = 0.
 \eeq
This means that  $F: (u_{-}, u_{+}) \rightarrow  \R^N$
is a null-geodesic map for the minisupermetric ${\cal G}$.
Thus, we are led to the Lagrange system (\ref{3.2.12})
with the minisupermetric  ${\cal G}$ defined in (\ref{3.2.3n}).

The problem of integrability will be simplified if we integrate
the Lagrange equations corresponding to $\Phi^s$:
 \bear{3.2.14}
  \frac d{du}\left(\exp(-2U^s(\sigma))\dot\Phi^s\right)=0
  \Longleftrightarrow
  \dot\Phi^s=Q_s \exp(2U^s(\sigma)),
 \ear
where $Q_s$ are constants, $s \in S$.
Here $(F^{\hat A})= (\sigma^A, \Phi^s)$.
We put $Q_s\ne0$ for all  $s \in S$.

For fixed $Q=(Q_s,s\in S)$ the Lagrange equations for the Lagrangian
 (\ref{3.2.12})  corresponding to $(\sigma^A)=(\phi^i,\varphi^\alpha)$,
when equations (\ref{3.2.14}) are substituted, are equivalent to the
Lagrange equations for the Lagrangian
 \beq{3.2.16}
  L_Q=\frac12\hat G_{AB}\dot \sigma^A\dot \sigma^B-V_Q,
 \eeq
where
 \beq{3.2.17}
  V_Q=\frac12  \sum_{s\in S}  \eps_s Q_s^2 \exp[2U^s(\sigma)],
 \eeq
the matrix $(\hat G_{AB})$ is defined in (\ref{2.2.9}). The zero-energy
constraint (\ref{3.2.13}) reads
 \beq{3.2.18}
  E_Q= \frac12 \hat G_{AB}\dot \sigma^A \dot \sigma^B+ V_Q =0.
 \eeq

\subsubsection{ Toda-type solutions}

Here we are interested in exact solutions for a case
when $K_s =(U^s,U^s)\neq 0$,
for all $s\in S$, and the quasi-Cartan matrix
 \ber{3.1.2.2}
  A_{ss'} \equiv \frac{2(U^s,U^{s'})}{(U^{s'},U^{s'})},
 \eer
is a non-degenerate one. Here  some ordering in $S$ is assumed.

It follows from the non-degeneracy of the matrix
 (\ref{3.1.2.2})
that the vectors $U^s, s \in S,$ are  linearly
independent and, hence,  $|S| \leq n+ l$.

The  exact solutions were obtained in \cite{IK}:
 \bear{3.2.63}
  g= \biggl(\prod_{s \in S} f_s^{2d(I_s)h_s/(D-2)}\biggr)
  \biggl\{ \exp(2c^0 H+ 2\bar  c^0) \hat{g}^0  \\ \nn
  + \sum_{i =1}^{n} \Bigl(\prod_{s \in S }
  f_s^{- 2h_s \delta_{i I_s} }\Bigr)
  \exp(2c^i H+ 2\bar  c^i) \hat{g}^i \biggr\},
  \\  \label{3.2.64}
  \exp(\varphi^\alpha) =
  \left( \prod_{s\in S} f_s^{h_s \chi_s\lambda_{a_s}^\alpha} \right)
  \exp(c^\alpha H +\bar c^\alpha),
 \ear
 $\alpha=1,\dots,l$ and $F^a=\sum_{s\in S}{\cal F}^s\delta_{a_s}^a$
with
 \bear{3.2.57}
  {\cal F}^s= Q_s
  \left( \prod_{s' \in S}  f_{s'}^{- A_{s s'}} \right) dH \wedge\tau(I_s),
  \qquad s\in S_e, \\
  \label{3.2.58}
  {\cal F}^s
  =  Q_s (*_0 d H) \wedge \tau(\bar I_s),
   \qquad s\in S_m,
 \ear
where  $*_0 = *[g^0]$ is the Hodge operator on $(M_0,g^0)$.
Here
 \beq{3.2.65}
   f_s = f_s(H) = \exp(- q^s(H)),
 \eeq
where $q^s(u)$ is a solution to  Toda-like equations
 \beq{3.2.35}
  \ddot{q^s} = -  B_s \exp( \sum_{s' \in S} A_{s s'} q^{s'} ),
 \eeq
with $ B_s =  K_s \eps_s Q_s^2$,
 $s \in S$, and $H = H(x)$ ($x \in M_0$) is a harmonic function on
 $(M_0,g^0)$. Vectors $c=(c^A)$ and $\bar c=(\bar c^A)$ satisfy the linear
constraints
 \bear{3.2.47}
  U^s(c)=
  \sum_{i \in I_s}d_ic^i-\chi_s\lambda_{a_s\alpha}c^\alpha=0,
  \quad
  U^s(\bar c)= 0,
 \ear
 $s\in S$, and
 \beq{3.2.52a}
  c^0 = \frac1{2-d_0}\sum_{j=1}^n d_j c^j,
  \quad  \bar  c^0 = \frac1{2-d_0}\sum_{j=1}^n d_j \bar c^j.
 \eeq
The zero-energy constraint reads
 \bear{3.2.53}
  2E_{TL} + h_{\alpha\beta}c^\alpha c^\beta+ \sum_{i=1}^n d_i(c^i)^2+
  \frac1{d_0-2}\left(\sum_{i=1}^nd_ic^i\right)^2 = 0,
 \ear
where
 \beq{3.2.54a}
  E_{TL} = \frac{1}{4}  \sum_{s,s' \in S} h_s
  A_{s s'} \dot{q^s} \dot{q^{s'}}
  + \sum_{s \in S} A_s  \exp( \sum_{s' \in S} A_{s s'} q^{s'} ),
  \eeq
is an integration constant (energy) for the solutions from
(\ref{3.2.35}) and $A_s =  \frac12  \eps_s Q_s^2$.

We note that the equations
 (\ref{3.2.35}) correspond to the Lagrangian
 \beq{3.2.37}
  L_{TL} = \frac{1}{4}  \sum_{s,s' \in S} h_s  A_{s s'} \dot{q^s}
  \dot{q^{s'}}
  - \sum_{s \in S} A_s  \exp( \sum_{s' \in S} A_{s s'} q^{s'} ),
  \eeq
where $h_s = K_s^{-1}$.

Thus, the solution is presented
by relations  (\ref{3.2.63})-(\ref{3.2.65})
with the functions  $q^s$  defined in
 (\ref{3.2.35}) and the relations on the parameters of
solutions $c^A$, $\bar c^A$ $(A= i,\alpha,0)$,
imposed in  (\ref{3.2.47}),
 (\ref{3.2.52a}) and  (\ref{3.2.53}).

\section{Classical and quantum cosmological-type solutions}

Here we consider the case $d_0 =1$, $M_0 = \R$, i.e.
we are interesting in applications to
the sector with one-variable dependence.
We consider the manifold
 \beq{4.1}
  M = (u_{-},u_{+})  \times M_{1} \times \ldots \times M_{n}
 \eeq
with the metric
 \beq{4.2}
  g= w \e^{2{\gamma}(u)} du \otimes du +
  \sum_{i=1}^{n} \e^{2\phi^i(u)} {\hat g}^i ,
 \eeq
where $w=\pm 1$, $u$ is a distinguished coordinate which, by
convention, will be called ``time'';
 $(M_i,g^i)$ are oriented and connected Einstein spaces
(see (\ref{2.13})), $i=1,\dots,n$.
The functions $\gamma,\phi^i$: $(u_-,u_+)\to \R$ are smooth.

Here we adopt the p-brane ansatz from Sect. 2.
putting $g^0= w  du \otimes du$.

 \subsection{ Lagrange dynamics}

It follows from sect. 2.2 that the
equations of motion  and the Bianchi
identities for the field configuration under consideration
(with the restrictions from subsect. 2.2.1 imposed)
are equivalent to equations of motion for
 1-dimensional $\sigma$-model with the action
  \beq{4.1.1}
   S_{\sigma} = \frac{\mu}2
   \int du {\cal N} \biggl\{G_{ij}\dot\phi^i\dot\phi^j
   +h_{\alpha\beta}\dot\varphi^{\alpha}\dot\varphi^{\beta}
   +\sum_{s\in S}\eps_s\exp[-2U^s(\phi,\varphi)](\dot\Phi^s)^2
   -2{\cal N}^{-2}V_{w}(\phi)\biggr\},
  \eeq
where $\dot x\equiv dx/du$,
 \beq{4.1.2}
   V_{w} = -w V = -w\Lambda\e^{2\gamma_0(\phi)}+
  \frac w2\sum_{i =1}^{n} \xi_i d_i \e^{-2 \phi^i + 2 {\gamma_0}(\phi)}
 \eeq
is the potential with
 $\gamma_0(\phi)  \equiv \sum_{i=1}^nd_i\phi^i$,
and
 ${\cal N}=\exp(\gamma_0-\gamma)>0$
is the lapse function, $U^s = U^s(\phi,\varphi)$ are defined
in (\ref{2.2.11}), $\eps_s$ are defined in (\ref{2.2.13})
for $s=(a_s,v_s,I_s)\in S$, and $G_{ij}=d_i\delta_{ij}-d_id_j$
are components of the "pure cosmological" minisupermetric, $i,j=1,\dots,n$
\cite{IMZ}.

In the electric case $({\cal F}^{(a,m,I)}=0)$ for finite internal space
volumes $V_i$ the action (\ref{4.1.1}) coincides with the
action (\ref{2.1}) if
 $\mu=-w/\kappa_0^2$, $\kappa^{2} = \kappa^{2}_0 V_1 \ldots V_n$.

Action (\ref{4.1.1}) may be also written in the form
 \beq{4.1.6}
   S_\sigma=\frac\mu2\int du{\cal N}\left\{
  {\cal G}_{\hat A\hat B}(X)\dot X^{\hat A}\dot X^{\hat B}-
  2{\cal N}^{-2}V_w \right\},
 \eeq
where $X = (X^{\hat A})=(\phi^i,\varphi^\alpha,\Phi^s)\in
 {\R}^{N}$, $N = n +l + |S|$, and minisupermetric
 ${\cal G}$ is defined in (\ref{3.2.3n}).

 {\bf Scalar products.}
The minisuperspace metric (\ref{3.2.3n}) may be also written in the form
 ${\cal G}=\hat G+\sum_{s\in S}\eps_s
  \e^{-2U^s(\sigma)}d\Phi^s\otimes d\Phi^s$,
where $\sigma = (\sigma^A) = (\phi^i,\varphi^\alpha)$,
 \bear{4.1.8}
   \hat G=\hat G_{AB}d \sigma^A
   \otimes d \sigma^B=G_{ij}d\phi^i\otimes d\phi^j+
   h_{\alpha\beta}d\varphi^\alpha\otimes d\varphi^\beta,
 \ear
is truncated minisupermetric and $U^s(\sigma)=U_A^s \sigma^A$ is defined in
 (\ref{2.2.11}).
The potential (\ref{4.1.2})
reads
 \beq{4.1.9}
  V_w=(-w\Lambda)\e^{2U^\Lambda(\sigma)}+\sum_{j=1}^n\frac w2\xi_jd_j
  \e^{2U^j(\sigma)},
 \eeq
where
 \bear{4.1.10}
   U^j(\sigma)=U_A^j \sigma^A=-\phi^j+\gamma_0(\phi),
   \qquad (U_A^j)=(-\delta_i^j+d_i,0),
   \\ \label{4.1.11}
   U^\Lambda(\sigma)=U_A^\Lambda \sigma^A=\gamma_0(\phi),
   \qquad (U_A^\Lambda)=(d_i,0).
 \ear

The integrability of the Lagrange system (\ref{4.1.6}) crucially depends
upon the scalar products of co-vectors $U^\Lambda$, $U^j$, $U^s$
(see (\ref{3.1.1})).
These products are defined by
 (\ref{3.1.4}) and the following relations \cite{IMC}
 \bear{4.1.14}
    (U^i,U^j)=\frac{\delta_{ij}}{d_j}-1,
    \\
    \label{4.1.15}
    (U^i,U^\Lambda)=-1,
  \qquad
  (U^\Lambda,U^\Lambda)=-\frac{D-1}{D-2},
   \\ \label{4.1.17}
   (U^s,U^i)=-\delta_{iI_s}, \qquad
   (U^s,U^\Lambda)=\frac{d(I_s)}{2-D},
 \ear
where $s=(a_s,v_s,I_s) \in S$;
 $i,j= 1,\dots,n$.

 {\bf Toda-like representation.}
We put $\gamma= \gamma_0(\phi)$, i.e. the harmonic
time gauge is considered.  Integrating
the Lagrange equations corresponding to $\Phi^s$
(see (\ref{3.2.14})) we are led to the Lagrangian
from (\ref{3.2.16}) and the zero-energy constraint (\ref{3.2.18})
with the modified potential
 \beq{4.1.19}
   V_Q=V_w +\frac12\sum_{s\in S} \eps_sQ_s^2\exp[2U^s(\sigma)],
 \eeq
where $V_w$ is defined in  (\ref{4.1.2}).

\subsection{Classical solutions with $\Lambda = 0$}

Here we consider classical solutions with $\Lambda = 0$.

\subsubsection{Solutions with Ricci-flat spaces}

Let all spaces be Ricci-flat, i.e. $ \xi_1 =\dots=\xi_n=0$.

Since  $H(u)= u$ is a harmonic function on $(M_0,g^0)$
with $g^0 = w du\otimes du$  we get
for the metric and scalar fields from
 (\ref{3.2.63}), (\ref{3.2.64}) \cite{IK}
 \bear{4.3.3}
   g= \biggl(\prod_{s \in S} f_s^{2d(I_s)h_s/(D-2)}\biggr)
   \biggl\{ \exp(2c^0 u+ 2\bar  c^0) w du \otimes du   \\ \nonumber
   + \sum_{i =1}^{n} \Bigl(\prod_{s\in S }
   f_s^{- 2h_s \delta_{i I_s} }\Bigr)
   \exp(2c^i u + 2\bar  c^i) {\hat g}^i \biggr\},
   \\  \label{4.3.4}
   \exp(\varphi^\alpha) = \left( \prod_{s\in S}
   f_s^{h_s \chi_s\lambda_{a_s}^\alpha} \right)
   \exp(c^\alpha u +\bar c^\alpha),
 \ear
 $\alpha=1,\dots,l$, where   $f_s = f_s(u) = \exp(- q^s(u))$ and
 $q^s(u)$ obey  Toda-like equations (\ref{3.2.35}).

Relations (\ref{3.2.52a}) and  (\ref{3.2.53}) take the form
 \bear{4.3.5}
   c^0 = \sum_{j=1}^n d_j c^j,
 \qquad  \bar  c^0 = \sum_{j=1}^n d_j \bar c^j,
  \\  \label{4.3.6c}
  2E_{TL} + h_{\alpha\beta}c^\alpha c^\beta+ \sum_{i=1}^n d_i(c^i)^2
 - \left(\sum_{i=1}^nd_ic^i\right)^2 = 0,
 \ear
with $E_{TL}$  from (\ref{3.2.54a}) and all other relations
(e.g. constraints (\ref{3.2.47}) and
relations  for forms (\ref{3.2.57}) and (\ref{3.2.58}) with $H = u$)
are unchanged.  In a special  ${\bf A_m}$ Toda chain case this solution
was considered previously in \cite{GM1}.

\subsubsection{Solutions with one curved space}

The cosmological solution with Ricci-flat spaces
may be also  modified to the following case:
 $ \xi_1 \ne0, \quad \xi_2=\ldots=\xi_n=0$,
i.e. one space is curved and others are Ricci-flat and
 $1 \notin I_s$,
 $s  \in S$,
i.e. all ``brane'' submanifolds  do not  contain $M_1$.

The potential $(\ref{3.2.17})$ is modified for $\xi_1 \ne0$
as follows (see \rf{4.1.19})
 \beq{4.3.6}
   V_Q=\frac12  \sum_{s\in S}  \eps_s Q_s^2 \exp[2U^s(\sigma)]
   + \frac12 w\xi_1 d_1  \exp[2U^1(\sigma)],
 \eeq
where $U^1(\sigma)$ is defined in  (\ref{4.1.10}) ($d_1 > 1$).

For the scalar products we get from (\ref{4.1.14}) and (\ref{4.1.17})
 \bear{4.3.7}
  (U^1,U^1)=\frac1{d_1}-1<0, \qquad (U^1,U^{s})=0
 \ear
for all $s\in S$. The solution in the case under consideration
may be obtained   by a little modification
of the solution from the previous section
(using (\ref{4.3.7}) and relations
 $U^{1i}= - \delta_1^i/d_1$,  $U^{1\alpha}=0$).
We get  \cite{IK}
 \bear{4.3.19}
  g= \biggl(\prod_{s \in S} [f_s(u)]^{2 d(I_s) h_s/(D-2)} \biggr)
  \biggl\{[f_1(u)]^{2d_1/(1-d_1)}\exp(2c^1u + 2 \bar c^1)\\ \nn
  \times[w du \otimes du+ f_1^2(u) \hat{g}^1] +
  \sum_{i = 2}^{n} \Bigl(\prod_{s\in S}
  [f_s(u)]^{- 2 h_s  \delta_{i I_s} }
  \Bigr)\exp(2c^i u+ 2 \bar c^i) \hat{g}^i \biggr\}.
   \\ \label{4.3.19a}
  \exp(\varphi^\alpha) =
  \left( \prod_{s\in S} f_s^{h_s \chi_s \lambda_{a_s}^\alpha} \right)
  \exp(c^\alpha u + \bar c^\alpha),
 \ear
and
 $F^a= \sum_{s \in S} \delta^a_{a_s} {\cal F}^{s}$
with forms
 \bear{4.3.19c}
   {\cal F}^s= Q_s
  \left( \prod_{s' \in S}  f_{s'}^{- A_{s s'}} \right) du \wedge\tau(I_s),
  \qquad s\in S_e,  \\
  \label{4.3.19d}
  {\cal F}^s= Q_s \tau(\bar I_s), \qquad s \in S_m
 \ear
 $Q_s \neq 0$, $s \in S$.
Here  $f_s = f_s(u) = \exp(- q^s(u))$ where
 $q^s(u)$ obeys  Toda-like equations (\ref{3.2.35})
and
 \bear{4.3.10}
  f_1(u) =R \sinh(\sqrt{C_1}(u-u_1)), \ C_1>0, \ \xi_1 w>0;
  \\ \label{4.3.11}
  R \sin(\sqrt{|C_1|}(u-u_1)), \ C_1<0, \  \xi_1 w>0; \\ \label{4.3.12}
  R \cosh(\sqrt{C_1}(u-u_1)),  \ C_1>0, \ \xi_1w <0; \\ \label{4.3.13}
   \left|\xi_1(d_1-1)\right|^{1/2}        , \ C_1=0,  \ \xi_1w>0,
 \ear
 $u_1, C_1$ are constants and $R =  |\xi_1(d_1-1)/C_1|^{1/2}$.

Vectors $c=(c^A)$ and $\bar c=(\bar c^A)$ satisfy the linear constraints
 \bear{4.3.15}
   U^r(c)= U^r(\bar c)= 0, \qquad r = s,1,
 \ear
(for $r =s$ see (\ref{3.2.47}))
and the zero-energy constraint
 \beq{4.3.17}
   C_1\frac{d_1}{d_1-1}= 2 E_{TL} +
   h_{\alpha\beta}c^\alpha c^\beta+ \sum_{i=2}^nd_i(c^i)^2+
  \frac1{d_1-1}\left(\sum_{i=2}^nd_ic^i\right)^2.
 \eeq

 {\bf Restriction 1} ( see subsect. 2.2.1) forbids certain intersections of
two $p$-branes with the same color index for  $n_1 \geq 2$.
 {\bf Restriction 2} is  satisfied identically in this case.

This solution in a special case
of ${\bf A_m}$ Toda chain was obtained earlier  in \cite{GM1}
(see also  \cite{LMMP}). For special sets of parameters
one can get so-called $S$-brane and flux-brane solutions
\cite{I-s,I-f} (for pioneering flux-brane solution see also \cite{GR}.)

 \subsubsection{Block-orthogonal solutions}

Let  us consider block-orthogonal case:
 (\ref{3.1.5}), (\ref{3.1.6}). In this case
 we get $f_s = \bar{f}_s^{b_s}$
where $b_s = 2 \sum_{s' \in S} A^{s s'}$,
  $(A^{ss'})= (A_{ss'})^{-1}$  and
 \bear{4.3.20n}
 \bar{f}_s(u)=
  R_s \sinh(\sqrt{C_s}(u-u_s)), \;
   C_s>0, \; \eta_s\eps_s<0; \\ \label{4.3.22n}
  R_s \sin(\sqrt{|C_s|}(u-u_s)), \;
   C_s<0, \; \eta_s\eps_s<0; \\ \label{4.3.23n}
  R_s \cosh(\sqrt{C_s}(u-u_s)), \;
   C_s>0, \; \eta_s\eps_s>0; \\ \label{4.3.24n}
   \frac{|Q^s|}{|\nu_s|}(u-u_s), \; C_s=0, \; \eta_s\eps_s<0,
 \ear
where $R_s = |Q_s|/(|\nu_s||C_s|^{1/2})$,
 $\eta_s \nu_s^2 = b_s h_s$, $\eta_s= \pm 1$,
 $C_s$, $u_s$ are constants, $s \in S$.
The constants $C_s$, $u_s$ are coinciding inside blocks:
 $u_s = u_{s'}$, $C_s = C_{s'}$,
 $s,s' \in S_i$, $i = 1, \ldots, k$.
The ratios $\eps_s Q_s^2/(b_s h_s)$ are
also coinciding inside blocks, or, equivalently,
 \bear{4.3.25n}
  \frac{\eps_s Q_s^2}{b_s h_s} = \frac{\eps_{s'} Q_{s'}^2}{b_{s'} h_{s'}},
 \ear
 $s,s' \in S_i$, $i = 1, \ldots, k$.

Here
 \beq{aa}
   E_{TL} = \sum_{s \in S} C_s \eta_s \nu_s^2.
 \eeq

The solution (\ref{4.3.19})-(\ref{4.3.19d}) with block-orthogonal
set of vectors was obtained in \cite{IMJ1} (for non-composite case see
also \cite{Br1}). In the special  orthogonal case when: $|S_1| = \ldots =
 |S_k| = 1$, the solution  was obtained in \cite{IMJ}.

\subsection{Quantum solutions}

\subsubsection{ Wheeler--De Witt equation}

Here we fix the gauge as follows
 \beq{4.2.1}
  \gamma_0-\gamma=f(X),  \quad  {\cal N} = e^f,
 \eeq
where $f$: ${\cal M}\to{\bf R}$ is a smooth function. Then we obtain the
Lagrange system with the Lagrangian
 \beq{4.2.3}
   L_f=\frac\mu2\e^f{\cal G}_{\hat A\hat B}(X)
   \dot X^{\hat A}\dot X^{\hat B}-\mu\e^{-f}V_w
 \eeq
and the energy constraint
 \beq{4.2.4}
  E_f=\frac\mu2\e^f{\cal G}_{\hat A\hat B}(X)
  \dot X^{\hat A}\dot X^{\hat B}+\mu\e^{-f}V_w=0.
 \eeq

Using the standard prescriptions of (covariant and conformally
covariant)  quantization (see, for example,
 \cite{IMZ}) we are led to the Wheeler-DeWitt (WDW) equation
 \beq{4.2.5}
   \hat{H}^f \Psi^f \equiv
    \left(-\frac{1}{2\mu}\Delta\left[e^f{\cal G}\right]+
    \frac{a}{\mu}R\left[e^f{\cal G}\right]
    +e^{-f}\mu V_w \right)\Psi^f=0,
  \eeq
where $ a=a_N= (N-2)/8(N-1)$ and   $N = n+l +|S|$.

Here $\Psi^f=\Psi^f(X)$ is the so-called ``wave function of the universe''
corresponding to the $f$-gauge (\ref{4.2.1}) and satisfying the relation
 \beq{4.2.7}
  \Psi^f= e^{bf} \Psi^{f=0}, \quad b = (2-N)/2,
 \eeq
 ($\Delta[{\cal G}_1]$ and
  $R[{\cal G}_1]$ denote the Laplace-Beltrami operator and the scalar
curvature corresponding to ${\cal G}_1$, respectively).

{\bf Harmonic-time gauge} The WDW equation (\ref{4.2.5}) for $f=0$
 \beq{4.2.10}
  \hat H\Psi \equiv \left(-\frac{1}{2\mu}\Delta[{\cal G}]+
  \frac{a}{\mu}R[{\cal G}]+\mu V_w \right)\Psi=0,
  \eeq
where
 \beq{4.2.8}
   R[{\cal G}]=-\sum_{s\in S}(U^s,U^s)-
   \sum_{s,s'\in S}(U^s,U^{s'}).
  \eeq
and
 \beq{4.2.9}
    \tri[{\cal G}]
    =\e^{U(\sigma)}\frac\partial{\partial \sigma^A}\left(\hat G^{AB}
    \e^{-U(\sigma)}\frac\partial{\partial \sigma^B}\right)
    +\sum_{s\in S}\eps_s\e^{2U^s(\sigma)}
    \left(\frac\partial{\partial\Phi^s}\right)^2,
  \eeq
with  $U(\sigma)=\sum_{s\in S}U^{s}(\sigma)$.

 \subsubsection{Solutions with one curved factor space and
                orthogonal $U^s$}

Here as in subsect. 4.2.2  we put $\Lambda=0$,
 $\xi_1 \ne0, \quad \xi_2=\ldots=\xi_n=0$,
and $1 \notin I_s$,   $s  \in S$, i.e.
the space $M_1$  is curved and others are Ricci-flat and
all ``brane'' submanifolds  do not  contain $M_1$.
We also put orthogonality restriction on the vectors $U^s$:
 $(U^s,U^{s'})= 0$ for $s \neq s'$ and $K_s = (U^s,U^s) \neq 0$
for all $s \in S$.
In this case the  potential (\ref{4.1.9}) reads
 $V_w= \frac12w \xi_1d_1\e^{2U^1(\sigma)}$.
The truncated minisuperspace metric
(\ref{4.1.8}) may be diagonalized
by the linear transformation
 $z^A=S^A{}_B \sigma^B$, $(z^A)=(z^1,z^a,z^s)$
as follows
 \beq{4.2.24}
   \hat G=-dz^1\otimes dz^1+
 \sum_{s\in S}\eta_sdz^s\otimes dz^s+dz^a\otimes dz^b\eta_{ab},
 \eeq
where $a,b=2,\dots,n +l-|S|$; $\eta_{ab} =\eta_{aa} \delta_{ab};
\eta_{aa}= \pm 1$, $\eta_s = {\rm sign}(U^s,U^s)$ and
 $q_1 z^1=U^1(\sigma)$,
 $q_1\equiv\sqrt{|(U^1,U^1)|} = \sqrt{1 - d_1^{-1}}$,
 $q_sz^s=U^s(\sigma)$,
 $q_s=\nu_s^{-1} \equiv \sqrt{|(U^s,U^s)|}$
 $s=(a_s,v_s,I_s) \in S$.

We are seeking the solution to WDW equation (\ref{4.2.10}) by the method
of the separation of variables, i.e. we put
 \beq{4.2.30}
  \Psi_*(z)=\Psi_1 (z^1)\left(\prod_{s\in S}\Psi_s(z^s)\right)
  \e^{\im P_s\Phi^s}\e^{\im p_az^a}.
 \eeq
It follows from  the relation for the Laplace operator  (\ref{4.2.9})
that $\Psi_*(z)$ satisfies WDW equation
 (\ref{4.2.10}) if
 \bear{4.2.31}
   \left\{\left(\frac\partial{\partial z^1}\right)^2
  +\mu^2 w\xi_1d_1\e^{2q_1z^1}\right\}\Psi_1=2{\cal E}_1\Psi_1; \\
  \label{4.2.32}
  \left\{-\eta_s\e^{q_sz^s}\frac\partial{\partial z^s}
  \left(\e^{-q_sz^s}\frac\partial{\partial z^s}\right)+
  \eps_sP_s^2\e^{2q_sz^s}\right\}\Psi_s=2{\cal E}_s\Psi_s,
  \ear
 $s\in S$, and
 \beq{4.2.33}
  2{\cal E}_1+\eta^{ab}p_ap_b+2\sum_{s\in S}{\cal E}_s+
   2aR[{\cal G}]=0,
 \eeq
where $a = a_N= (N-2)/8(N-1)$ and
 $R[{\cal G}]=-2\sum_{s\in S}(U^s,U^s)=
  -2\sum_{s\in S}\eta_sq_s^2$.

 We obtain the following linearly independent
solutions to (\ref{4.2.31}) and (\ref{4.2.32}), respectively
 \bear{4.2.24n}
  \Psi_1(z^1)=B_{\omega_1}^1
  \left(   \sqrt{-w \mu^2 \xi_1d_1}\frac{\e^{q_1z^1}}{q_1}\right), \\
  \label{4.2.25n}
  \Psi_s(z^s)=\e^{q_sz^s/2}B_{\omega_s}^s
  \left(\sqrt{\eta_s\eps_sP_s^2}\frac{\e^{q_sz^s}}{q_s}\right),
 \ear
where
 $\omega_1=\sqrt{2{\cal E}_1}/q_1$,
 $\omega_s=\sqrt{\frac14- 2\eta_s{\cal E}_s\nu_s^2}$,
 $s\in S$, and $B_\omega^1,B_\omega^s=I_\omega,K_\omega$
are the modified Bessel function.

The general solution to the WDW equation (\ref{4.2.10}) is a superposition
of the "separated" solutions (\ref{4.2.30}):
 \beq{4.2.28a}
  \Psi(z)=\sum_B\int dpdPd{\cal E} C(p,P,{\cal E},B)
  \Psi_*(z|p,P,{\cal E},B),
 \eeq
where $p=(p_a)$, $P=(P_s)$, ${\cal E}=({\cal E}_s, {\cal E}_1)$,
  $B=(B^1,B^s)$,  $B^1,B^s=I,K$; and
  $\Psi_*=\Psi_*(z|p,P,{\cal E},B)$ is given by relation
 (\ref{4.2.30}), (\ref{4.2.24n})--(\ref{4.2.25n}) with ${\cal E}_1$ from
 (\ref{4.2.33}). Here $C(p,P,{\cal E},B)$ are smooth enough functions.

{\bf Solution with  $n$ Ricci-flat spaces.} In the case $\xi_1 = 0$
the solution to WDW equation is given by the following
 modification of the relation (\ref{4.2.30}) and
(\ref{4.2.33}), respectively
 \bear{4.2.30a}
  \Psi_*(z)= \left(\prod_{s\in S}\Psi_s(z^s)\right)
  \e^{\im P_s\Phi^s}\e^{\im p_az^a}, \\
   \label{4.2.33a}
  \eta^{ab}p_ap_b+2\sum_{s\in S}{\cal E}_s+
   2aR[{\cal G}]=0,
 \ear
where here $a,b=1,\dots,n +l-|S|$. In this case
(\ref{4.2.24}) should be also modified
\beq{4.2.24a}
   \hat G=
 \sum_{s\in S}\eta_sdz^s\otimes dz^s+dz^a\otimes dz^b\eta_{ab},
 \eeq
(with $a,b=1,\dots,n +l-|S|$). Here restriction
$1 \notin I_s$, $s  \in S$, should be removed and
$z^1$ is not obviously related to $U^1$.

 \subsection{Spherically-symmetric solutions with a horizon}

Here we consider the spherically-symmetric case
of the metric (\ref{4.3.19}), i.e. we put
  $w = 1, \quad M_1 = S^{d_1}$,  $g^1 = d \Omega^2_{d_1}$,
where $d \Omega^2_{d_1}$ is the canonical metric on a unit sphere
  $S^{d_1}$, $d_1 \geq 2$. In this case $\xi^1 = d_1 -1$.
We put  $M_2 = \R$, $g^2 = - dt \otimes dt$,
i.e.  $M_2$ is a time manifold. We also assume that
  $K_s = (U^s,U^s) \neq 0$, $s \in S$, and
 \beq{5.2.2a}
   {\rm det}((U^s,U^{s'})) \neq 0.
 \eeq

We put $C_1 \geq 0$.
When the matrix $(h_{\alpha\beta})$ is positive
definite  and
 \beq{5.2.18}
  2 \in I_s, \quad \forall s \in S,
 \eeq
i. e. all p-branes have a common time direction $t$,
the horizon condition (of infinite time of light propagation)
singles out the unique
solution with $C_1 > 0$ and linear asymptotics at infinity
 $q^s = - \beta^s u + \bar \beta^s  + o(1)$,
 $u \to +\infty$, where $\beta^s, \bar \beta^s$ are
constants, $s \in S$,
 \cite{IMp2,IMp3}.

The  solutions with horizon
have the following form
 \cite{IMp1,IMp2,IMp3}
 \bear{5.2.30}
  g= \Bigl(\prod_{s \in S} H_s^{2 h_s d(I_s)/(D-2)} \Bigr)
    \biggl\{ \left(1 - \frac{2\mu}{R^{\bar{d}}}\right)^{-1} dR \otimes dR
   + R^2  d \Omega^2_{d_1}  \\ \nn
   - \Bigl(\prod_{s \in S} H_s^{-2 h_s} \Bigr)
   \left(1 - \frac{2\mu}{R^{\bar{d}}}\right) dt \otimes dt
   + \sum_{i = 3}^{n} \Bigl(\prod_{s \in S}
   H_s^{-2 h_s \delta_{iI_s}} \Bigr) \hat{g}^i  \biggr\},
   \\  \label{5.2.31}
   \exp(\varphi^\alpha)=
   \prod_{s\in S} H_s^{h_s \chi_s \lambda_{a_s}^\alpha},
 \ear
where $F^a= \sum_{s \in S} \delta^a_{a_s} {\cal F}^{s}$,
and
 \beq{5.2.32}
   {\cal F}^s= - \frac{Q_s}{R^{d_1}}
  \left( \prod_{s' \in S}  H_{s'}^{- A_{s s'}} \right) dR \wedge\tau(I_s),
 \eeq
 $s\in S_e$,
 \beq{5.2.33}
  {\cal F}^s= Q_s \tau(\bar I_s),
 \eeq
 $s\in S_m$.
Here $Q_s \neq 0$, $h_s =K_s^{-1}$, $s \in S$, and the
quasi-Cartan matrix $(A_{s s'})$ is non-degenerate.

Functions $H_s$ obey the following
differential equations
 \beq{5.3.1}
  \frac{d}{dz} \left( \frac{(1 - 2\mu z)}{H_s}
  \frac{d}{dz} H_s \right) = \bar B_s
  \prod_{s' \in S}  H_{s'}^{- A_{s s'}},
 \eeq
where $H_s(z) > 0$, $\mu > 0$,
 $z = R^{-\bar d} \in (0, (2\mu)^{-1})$ and \\
 $\bar B_s =  \eps_s K_s Q_s^2/ \bar{d}^2 \neq 0$
equipped with the boundary conditions
 \bear{5.3.2a}
    H_{s}((2\mu)^{-1} -0) = H_{s0} \in (0, + \infty), \\
   \label{5.3.2b}
    H_{s}(+ 0) = 1,
 \ear
 $s \in S$.

 Equations (\ref{5.3.1}) are equivalent to Toda-like
 equations. The first boundary condition
 leads to  a regular horizon at  $R^{\bar{d}} =  2 \mu$.
 The second condition  (\ref{5.3.2b})  guarantees the asymptotical flatness
 (for $R \to +\infty$) of the $(2+d_1)$-dimensional section of the metric.

There exist solutions to eqs. (\ref{5.3.1})-(\ref{5.3.2b})
of polynomial type. The simplest example occurs in orthogonal
case \cite{AIV,Oh,IMJ,BIM}:
 $(U^s,U^{s'})= 0$, for  $s \neq s'$, $s, s' \in S$. In this case
 $(A_{s s'}) = {\rm diag}(2,\ldots,2)$ is a Cartan matrix
for semisimple Lie algebra
 ${\bf A_1} \oplus  \ldots  \oplus  {\bf A_1}$
and
 \beq{5.3.5}
     H_{s}(z) = 1 + P_s z,
 \eeq
with $P_s \neq 0$, satisfying
 \beq{5.3.5a}
  P_s(P_s + 2\mu) = -\bar B_s,
 \eeq
 $s \in S$.

In \cite{Br1,IMJ2} this solution was generalized to a
block-orthogonal case  (\ref{3.1.5}), (\ref{3.1.6}).
In this case (\ref{5.3.5}) is modified as follows
 \beq{5.3.8}
  H_{s}(z) = (1 + P_s z)^{b_s},
 \eeq
where $b_s = 2 \sum_{s' \in S} A^{s s'}$
and parameters $P_s$  and are coinciding inside
blocks, i.e. $P_s = P_{s'}$ for $s, s' \in S_i$, $i =1,\dots,k$.
Parameters $P_s \neq 0 $ satisfy the relations
 \beq{5.3.5b}
   P_s(P_s + 2\mu) = - \bar B_s/b_s,
 \eeq
 $b_s \neq 0$,
and parameters $\bar B_s/b_s$  are also  coinciding inside
blocks.

{\bf Conjecture.} {\em Let $(A_{s s'})$ be  a Cartan matrix
for a  semisimple finite-dimensional Lie algebra $\cal G$.
Then  the solution to eqs. (\ref{5.3.1})-(\ref{5.3.2b})
(if exists) is a polynomial
\beq{5.3.12}
H_{s}(z) = 1 + \sum_{k = 1}^{n_s} P_s^{(k)} z^k,
\eeq
where $P_s^{(k)}$ are constants,
$k = 1,\ldots, n_s$; $n_s = b_s = 2 \sum_{s' \in S} A^{s s'} \in \N$
and $P_s^{(n_s)} \neq 0$,  $s \in S$}.

In this case all powers $n_s$
are  natural numbers  coinciding with the components
of twice the  dual Weyl vector in the basis of simple coroots
\cite{FS}.
In extremal case ($\mu = + 0$) an a analogue of this conjecture
was suggested previously in \cite{LMMP}.
Conjecture 1 was verified for ${\bf A_m}$ and ${\bf C_{m+1}}$ series of Lie
algebras in \cite{IMp2,IMp3}.
Explicit relations  for ${\bf C_2}$ and  ${\bf A_3}$ algebras were
obtained in \cite{GrIK} and \cite{GrIM}, respectively.

\section{Billiard representation near the singularity}

It was found in that \cite{IMb1}
the cosmological models with $p$-branes may have
a ``never ending'' oscillating behaviour near the cosmological
singularity as it takes place in Bianchi-IX model \cite{BLK}.
Remarkably, this oscillating behaviour may
be described using the so-called billiard representation
near the singularity. In  \cite{IMb1} the billiard representation for a
cosmological model with a set of electro-magnetic composite $p$-branes in
a theory with the action (\ref{2.1})  was
obtained (see also \cite{DH,DHN} and references therein).

In terms of the Kasner parameters $\alpha = (\alpha^{A}) =
(\alpha^{i}, \alpha^{\gamma})$, satisfying relations
 \beq{6.1}
  \sum_{i=1}^{n} d_i \alpha^i =
  \sum_{i=1}^{n} d_i (\alpha^i)^2 +
  \alpha^{\beta} \alpha^{\gamma} h_{\beta \gamma}= 1,
 \eeq
the existence of never ending oscillating  behaviour near the
singularity takes place if for any $\alpha$ there exists
 $s \in S$ such that $(U^s,U^s) > 0$ and
 \beq{1.17}
  U^s(\alpha) =  U_A^{s} \alpha^A = \sum_{i\in I_s}d_i\alpha^i
  -\chi_s\lambda_{a_s \gamma}\alpha^{\gamma} \leq 0
 \eeq
  \cite{IMb1}.
Thus, $U$-vectors play a key role in determination
of possible oscillating behaviour near the singularity.
In \cite{IMb1} the relations (\ref{1.17})
were also interpreted in terms of
illumination of a (Kasner) sphere by point-like sources.

  {\bf General ``collision law''.}
The set of Kasner parameters  $(\alpha^{'A})$
after the collision with the $s$-th wall ($s \in S$) is defined
by the Kasner set before the collision $(\alpha^{A})$ according
to the following formula
  \beq{gcl}
  \alpha^{'A} =
               \frac{\alpha^A - 2 U^s(\alpha) U^{sA}(U^s,U^s)^{-1}}
               {1 - 2 U^s(\alpha) (U^s,U^{\Lambda})(U^s,U^s)^{-1}},
  \eeq
where $U^{s A} = \bar{G}^{AB} U^s_B$,
     $U^s(\alpha) =  U_A^{s} \alpha^A$ and co-vector $U^{\Lambda}$
is defined in (\ref{4.1.11}).

The formula (\ref{gcl}) follows just from the reflection ``law''
 \beq{gcl1}
    v^{'A} =  v^A - 2 U^s(v) U^{sA}(U^s,U^s)^{-1},
   \eeq
  for Kasner free ``motion'':  $\sigma^A = v^A t + \sigma_0^A$,
  ($t$ is harmonic time) and the definition of Kasner
  parameters:  $\alpha^A =  v^A/ U^{\Lambda}(v)$.
  In the special case of one
  scalar field and  $1$-dimensional factor-spaces
  (i.e.  $l= d_i =1$) this formula was suggested earlier
  in \cite{DH}.

\section{Conclusions}

Here we considered  several rather general families of exact solutions
in multidimensional gravity with a set of scalar fields and fields
of forms. These solutions describe composite (non-localized)
electromagnetic $p$-branes defined on products of {\em Ricci-flat}
(or sometimes Einstein) spaces of {\em arbitrary signatures}.
The metrics are block-diagonal and
all scale factors, scalar fields and fields of forms depend
on points of some (mainly Ricci-flat) manifold $M_0$.
The solutions include those depending upon harmonic functions,
cosmological and spherically-symmetric solutions
(e.g. black-brane ones).
Our scheme is based on the sigma-model representation
obtained in \cite{IMC}.

Here we considered also the Wheeler-DeWitt  equation
for $p$-brane cosmology in d'Alembertian (covariant)
and conformally covariant form
and integrated it for orthogonal $U$-vectors (and $n -1$

Ricci-flat internal spaces).
We also overviewed general classes of "cosmological"
and spherically symmetric solutions governed by Toda-type
equations, e.g. black brane ones. An interesting
point here is the appearance of polynomials for black brane
solutions when  brane intersections are governed by Cartan
matrices of finite-dimensional simple Lie algebras
\cite{IMp1}. It should be noted
that post-Newtonian parameters corresponding to
certain 4-dimensional section of metrics were  calculated
in \cite{IMJ2, IMp2}.

Another topic of interest appearing here is the possible oscillating
behaviour in the models with $p$-branes, that
has a description by billiards in  hyperbolic
(Lobachevsky) spaces \cite{IMb1}.


\begin{center}
{\bf Acknowledgments}
\end{center}

This work was supported in part
by the DFG grant  436 RUS 113/236/O(R),
by the Russian Ministry of
Science and Technology and  Russian Foundation for Basic Research
grant. The author thanks Prof. V. de Sabbata and his colleagues
for kind hospitality during the School in Erice (in May 2003).


\end{document}